\documentclass[12pt]{aastex63}

\usepackage{graphicx}           
\usepackage{latexsym}           
\usepackage{amsmath}            
\usepackage{nccmath}            
\usepackage{bm}                 
\usepackage[english]{babel}
\usepackage{color}              
\usepackage{booktabs}           
\usepackage{multirow}           
\usepackage{natbib}

\newcommand{\del}[1]{\relax}%
\newcommand{\DELETED}[1]{\relax}%
\newcommand{\DEL}[1]{\relax}%
{\relax}%

\definecolor{violet}  {rgb}{1.0,0.0,1.0}

\definecolor{dviolet} {rgb}{0.75,0.0,1.0}

\definecolor{blue}    {rgb}{0.0,0.7,1.0}

\definecolor{lblue}   {rgb}{0.5,1,1}

\definecolor{dblue}   {rgb}{0.0,0.0,1.0}

\definecolor{blgr}    {rgb}{0.70,0.80,1.00}

\definecolor{navy}    {rgb}{0.00,0.00,0.48}

\definecolor{green}   {rgb}{0.7,1.0,0.0}

\definecolor{dgreen}  {rgb}{0.0,1.0,0.0}

\definecolor{lgreen}  {rgb}{0.0,0.8,0.0}

\definecolor{dg}      {rgb}{0.0,0.6,0.0}

\definecolor{orange}  {rgb}{1.0,0.5,0.0}

\definecolor{dorange} {rgb}{1.0,0.6,0.0}

\definecolor{brown}   {rgb}{0.1,0.1,0.0}

\definecolor{lbrown}  {rgb}{0.7,0.5,0.0}

\definecolor{red}     {rgb}{1,0.0,0.0}

\definecolor{dred}    {rgb}{0.6,0.0,0.0}

\definecolor{grey}    {rgb}{0.1,0.1,0.1}

\definecolor{lgrey}   {rgb}{0.5,0.5,0.5}

\definecolor{black}   {rgb}{0.0,0.0,0.0}

\newcommand\n            {\noindent}

\newcommand\m            {\medskip}
\newcommand\s            {\smallskip}

\newcommand\bn           {\bigskip\noindent}
\newcommand\mn           {\medskip\noindent}
\newcommand\sn           {\smallskip\noindent}
\newcommand\cl           {\centerline}

\newcommand\arcspt       {{$\buildrel{\prime\prime}\over .$}}
\newcommand\degree       {{\ifmmode^\circ\else$^\circ$\fi}} 
\newcommand\arcm         {{\ifmmode {'\ }\else$'     $\fi}} 
\newcommand\arcs         {{\ifmmode{''\ }\else$''    $\fi}} 

\newcommand{\bul}        {$\bullet$\ }
\newcommand\cge          {{$\gtrsim$}}
\newcommand\cle          {{$\lesssim$}}

\newcommand\eg           {{\it e.g.},}
\newcommand\ie           {{\it i.e.},}

\newcommand\JAB          {{$J_{AB}$} }

\newcommand\kms          {{km\ s$^{-1}$}}

\newcommand\iAB          {{$i_{\rm AB}$}}

\newcommand\MAB          {{$M_{\rm AB}$}}

\newcommand\mum          {{\micron}}

\newcommand\MUV          {{$M_{\rm UV}$} }

\newcommand\Ro           {{R$_{\odot}$}}
\newcommand\re           {{$r_{\rm e}$} }

\newcommand\rhl          {{$r_{\rm hl}$} }

\newcommand\Teff         {{$T_{\rm eff}$}}

\newcommand\vT           {{$v_{T}$}}

\defcitealias{driver_2016}{D16}
\defcitealias{windhorst_2011}{W11}

\def\ltsima{$\; \buildrel < \over \sim \;$}
\def\lsim{\lower.5ex\hbox{\ltsima}}
\def\gtsima{$\; \buildrel > \over \sim \;$}
\def\gsim{\lower.5ex\hbox{\gtsima}}

\newlength{\txw}
\newlength{\txh}

\begin{document}

\vspace*{-0.50cm}
\title{Galaxy Science with ORCAS: Faint Star-Forming Clumps to AB\cle 31 mag
and \re\cge 0\arcspt 01} 

\author{Rogier A. Windhorst}
\affiliation{School of Earth and Space Exploration, Arizona State University,
Tempe, AZ 85287-1404}

\author{Timothy Carleton} 
\affiliation{School of Earth and Space Exploration, Arizona State University,
Tempe, AZ 85287-1404}

\author{Seth H Cohen} 
\affiliation{School of Earth and Space Exploration, Arizona State University,
Tempe, AZ 85287-1404}

\author{Rolf Jansen} 
\affiliation{School of Earth and Space Exploration, Arizona State University,
Tempe, AZ 85287-1404}

\author{Rosalia O'Brien} 
\affiliation{School of Earth and Space Exploration, Arizona State University,
Tempe, AZ 85287-1404}

\author{Scott Tompkins} 
\affiliation{School of Earth and Space Exploration, Arizona State University,
Tempe, AZ 85287-1404}

\author{Daniel Coe}
\affiliation{Space Telescope Science Institute, 3700 San Martin Drive,
Baltimore, MD 21218}

\author{Jose M. Diego}
\affiliation{IFCA, Instituto de Fisica de Cantabria (UC-CSIC), Avenida de Los
Castros s/n, 39005 Santander, Spain}

\author{Brian Welch}
\affiliation{Center for Astrophysical Sciences, Department of Physics and
Astronomy, The Johns Hopkins University, 3400 N Charles St., Baltimore, MD
21218}

\email{Rogier.Windhorst@asu.edu}

\begin{abstract}
The NASA concept mission ORCAS (``Orbiting Configurable Artificial Star'') aims
to provide near diffraction-limited angular resolution at visible and
near-infra-red wavelengths using laser signals from space-based cubesats as
Adaptive Optics beacons for ground-based 8--30 meter telescopes, in particular
the 10 meter Keck Telescopes. When built as designed, ORCAS+Keck would deliver
images of $\sim$0\arcspt 01--0\arcspt 02 FWHM at 0.5--1.2 \mum\ wavelength that
reach AB\cle 31 mag for point sources in a few hours over a \cge
5$\times$5\arcs\ FOV that includes IFU capabilities. We summarize the potential
of high-resolution faint galaxy science with ORCAS. We show that the ability to
detect optical--near-IR point sources with \re \cge 0\arcspt 01 FWHM to AB\cle
31 mag will yield about 5.0$\times$10$^6$ faint star-forming (SF) clumps per
square degree, or $\sim$0.4 per arcsec$^2$, or one in every box of
1.6$\times$1.6\arcs. From recent HST lensing data, the typical {\it intrinsic
(\ie\ unlensed)} sizes of SF clumps at z$\simeq$1--7 will be \re$\simeq$1--80
m.a.s. to AB\cle 31 mag, with {\it intrinsic (\ie\ unmagnified)} fluxes as
faint as AB\cle 35--36 mag when searching with ORCAS around the critical curves
of the best lensing clusters imaged with HST and JWST. About half of these SF
clumps will have sizes below the ORCAS diffraction limit, and the other half
will be slightly resolved, but still mostly above the ORCAS surface brightness
(SB) limits. A \cge 5$\times$5\arcs\ ORCAS FOV may therefore provide just
enough compact SF clumps to do relative m.a.s.-astrometry. ORCAS will address
how galaxies assemble from smaller clumps to stable disks by measuring ages,
metallicities, and gradients of clumps within galaxies. Of particular interest
will be to follow up with ORCAS on caustic transits of individual stars in SF
clumps at z\cge 1--2 that have been detected with HST, and those that may be
detected with JWST at z$\simeq$6--17 at extreme magnifications ($\mu$\cge
10$^3$--10$^5$) for the first stars and their stellar mass black hole accretion
disks. The ability of ORCAS to monitor such stars for decades across the
(micro-)caustics provides a unique opportunity to obtain a statistical census
of individual stars at cosmological distances, leveraging the largest telescope
apertures that are available only on the ground.
\end{abstract}

\bn 

\n \section{Introduction}\label{sec1}

\sn In the last three decades, major progress has been made in studies of
galaxy assembly with the Hubble Space Telescope (HST) and through targeted
programs using Adaptive Optics (AO) on the world's best ground-based
facilities. It is not possible to review all these efforts here, and so we
refer the reader to more detailed reviews \citep[\eg][]{liv98, cris01, math06,
eller06, gardner06}. In \citet{windhorst_2008}, we reviewed the advantages of
high resolution science on high redshift galaxies from the ground as compared
to from space. In short, diffraction limited space-based imaging provides much
darker sky over a wider FOV, more stable PSF's, better dynamic range, and
therefore superior sensitivity, including in the vacuum-UV. But ground-based
multi-conjugate AO (MCAO) on 8-10 meter telescopes is {\it complementary} to
space-based imaging, as it can provide much higher spatial resolution --- and
spectral resolution --- than what space-based telescopes can currently do. 

One of the early discoveries by HST was that the numerous faint blue galaxies
are in majority late-type \citep[\eg][and references therein]{abr96, dri95,
dri98, gla95} and small \citep[][see Fig.\ref{fig:fig2} here]{ode96, coh03,
hathi_2008} star-forming objects. These are the building blocks of the giant
galaxies seen today \citep[\eg][]{pas96}. By measuring their distribution over
rest-frame type \citep{win02} versus redshift, HST has shown that galaxies of
all Hubble types formed over a wide range of cosmic time, but with a notable
transition around redshifts z$\simeq$0.5--1.0 \citep[\eg][]{dri95, dri98,
elm07}. This was done through HST programs like, \eg\ the Medium-Deep Survey
\citep{grif94}, the Hubble Deep Field \citep[HDF][]{williams_1996}, GOODS
\citep{giav04}, GEMS \citep{rix04}, the Hubble UltraDeep Field
\citep[HUDF][]{beckwith_2006}, COSMOS \citep{scoville_2007}, and CANDELS
\citep{grogin_2011, koekemoer_2011}. Coupled with models of galaxy formation,
these observations suggest that subgalactic units rapidly merged from the end
of reionization \citep{bou04a, yan04b} to grow bigger units at lower redshifts
\citep[\eg][]{pas96}. Merger products start to settle as galaxies with giant
bulges or large disks around redshifts z$\simeq$1 \citep[\eg][]{lil98, lil07}.
These have evolved mostly passively since then, resulting in giant galaxies
today \citep[\eg][]{dri98, coh03}. 

Star-forming clumps have also been studied at high resolution in lower redshift
turbulent galactic disks \citep[\eg][]{fisher_2017a, fisher_2017b}. The size
evolution of star-forming galaxies has been studied out to z\cle 7
\citep[\eg][]{ferg04, allen_2017}, where galaxy half-light or effective radii
\re approximately decrease with redshift as
$r_e$(z)$\propto$$r_e$(0)$\cdot$(1+z)$^{-s}$ with s$\simeq$0.9--1.2. This
strong size evolution reflects the hierarchical formation of galaxies, where
sub-galactic clumps and smaller galaxies merge over time to form the
larger/massive galaxies that we see today \citep[\eg][]{nav96}. It was the
reason that HST was so successful after its refurbishment in December 1993 at
identifying faint compact star-forming galaxies that form hierarchically at
z$\simeq$1--7 in the $\Lambda$CDM universe. The compact object sizes thereby
helped to mitigate the enormous (1+z)$^4$ cosmological SB-dimming that would
quickly render large extended objects undetectable at higher redshifts
\citep[\eg][]{windhorst_2008}. 

The combination of ORCAS (``Orbiting Configurable Artificial Star''; 
\url{https://asd.gsfc.nasa.gov/orcas/}) cube sat laser MCAO beacons with
ground-based 8--39 meter telescopes has the great potential to provide nearly
diffraction limited imaging over wider FOV's than possible with AO alone. For
instance, ORCAS combined with the 10 meter Keck telescope can provide PSF FWHM
values \cle 0\arcspt 01--0\arcspt 02 (10--20 mas) at 0.5--1.25 \mum\
wavelength, and still provide a sufficient FOV
(5$\times$5\arcs--10$\times$10\arcs) to detect a significant number of objects
to very faint fluxes (AB\cle 31 mag). In the optical--near-IR, ORCAS+Keck can
thus compete with space-based imaging in terms of increased spatial 
resolution, low sky-brightness in its very small pixels, and therefore increased
point source sensitivity. In the thermal infrared ($\lambda$\cge 2\mum), for
which JWST was designed and optimized \citep{gardner06}, space-based imaging
will remain superior in terms of PSF-stability, sky-brightness, depth, and FOV. 

\n \section{The Surface Density of Faint Star-Forming Clumps to AB\cle 31 mag
for ORCAS} \label{sec2} 

\begin{figure*}[!hptb]
\vspace*{-1.00cm}
\n \cl{
\hspace*{-0.00cm}
\includegraphics[width=0.5000\txw]{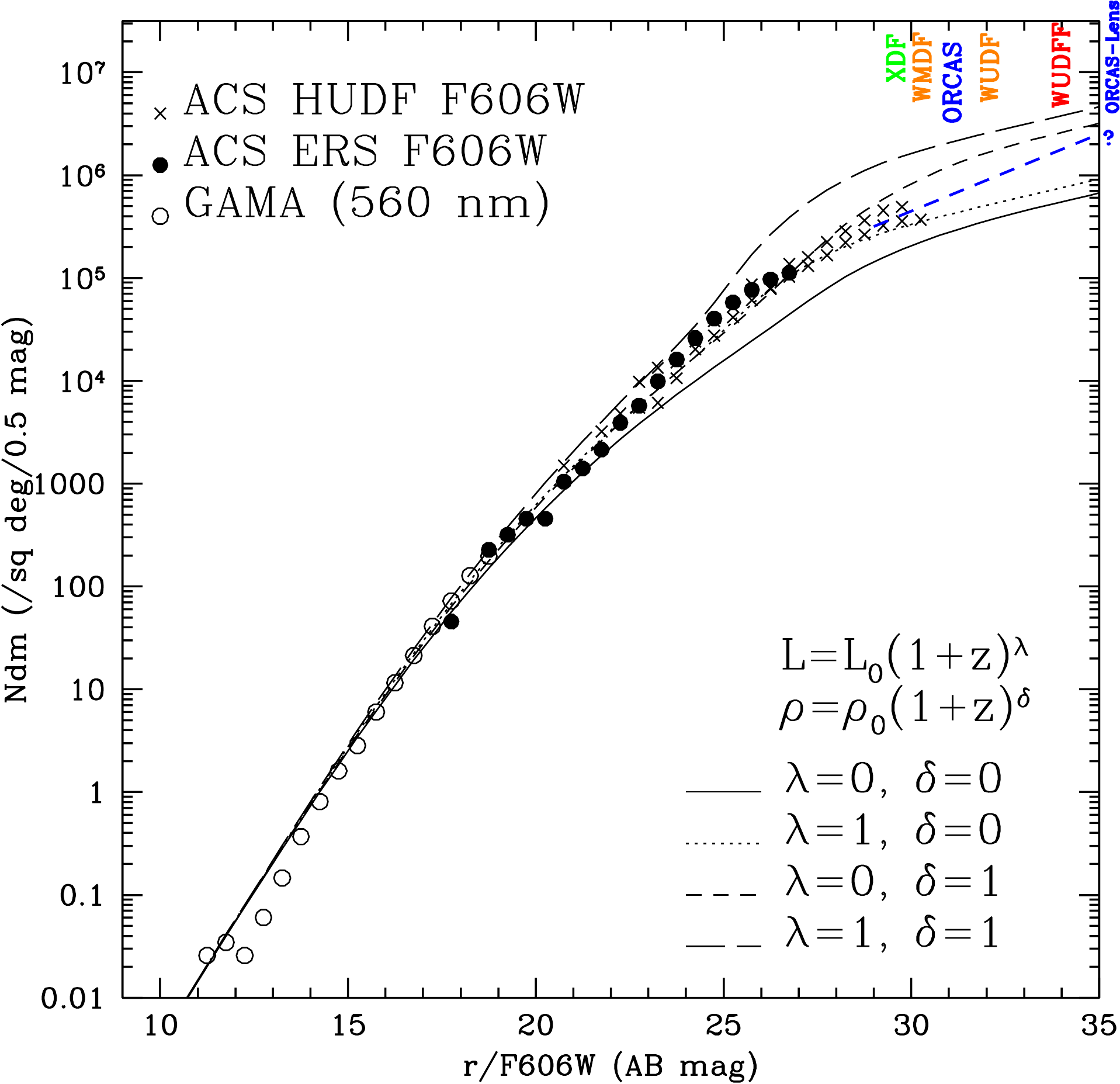}
\hspace*{+0.000cm}
\includegraphics[width=0.5000\txw]{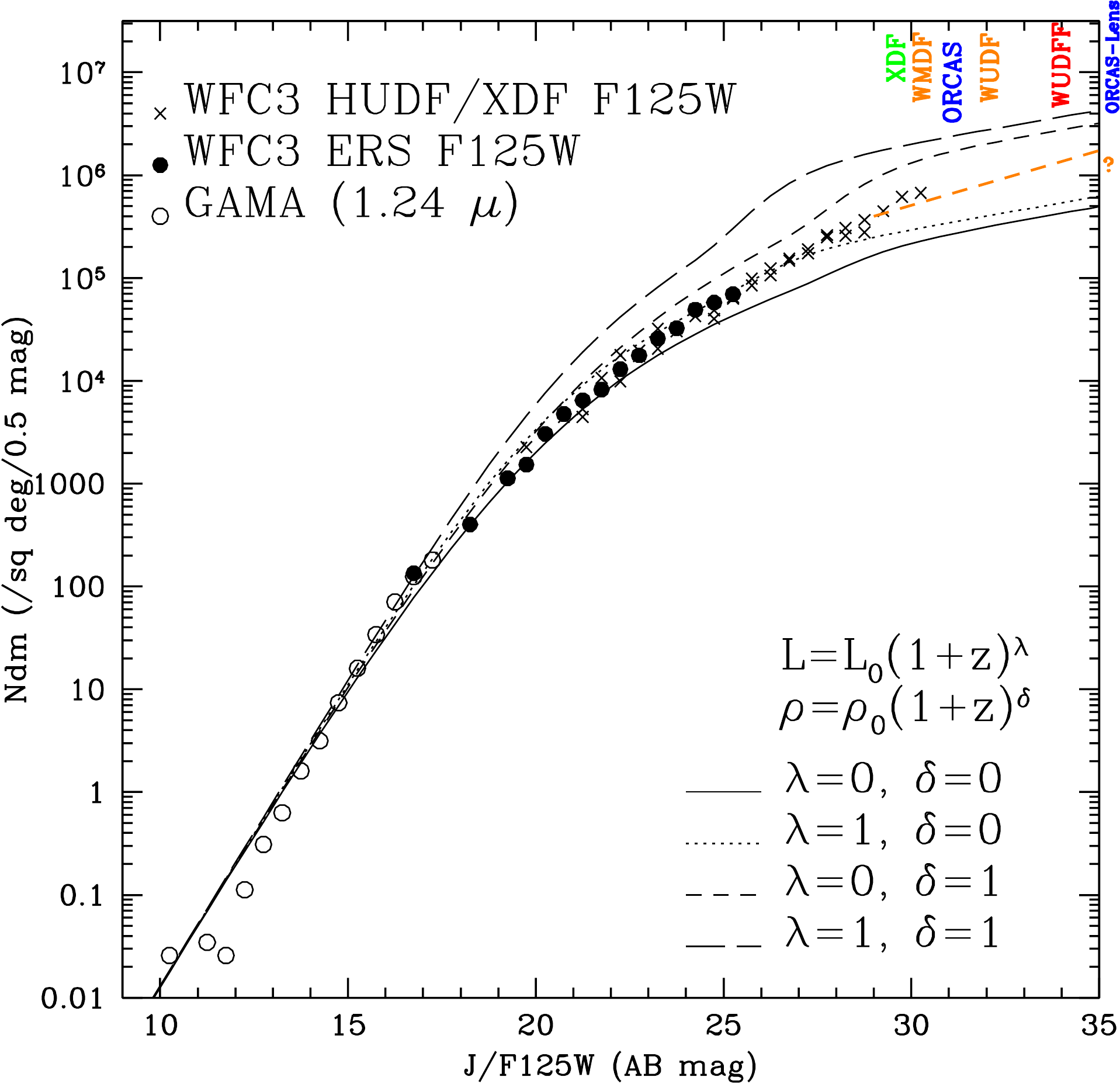}
}

\vspace*{-0.00cm}
\n \caption{
{\bf (a) [Left]:}\ Differential galaxy counts in the V-band or F606W filter. 
Data are from the ground-based GAMA survey and the HST ACS/WFC surveys in the 
WFC3 ERS and HUDF fields. Combined ground-based + HST-surveys cover 10\cle
AB\cle 30 mag \citep{windhorst_2011, driver_2016}. Simple luminosity+density
evolution models are shown and extrapolated to AB\cle 35 mag (see \S\
\ref{sec2} for details). 
{\bf (b) [Right]:}\ As Fig.~\ref{fig:fig1}a, but in the J-band or HST WFC3/IR
F125W filter. The colored labels indicate the various HUDF/XDF, Webb and ORCAS
detection limits without and with lensing. To AB\cle 31 mag in both filters,
ORCAS will yield about 5$\times$10$^6$ faint star-forming (SF) clumps per
square degree, or $\sim$0.4 per arcsec$^2$, or one in every box of
1.6$\times$1.6\arcs.}
\label{fig:fig1}
\end{figure*}

\begin{figure*}[!hptb]
\n\cl{
\vspace*{-1.00cm}
\includegraphics[width=0.880\txw]{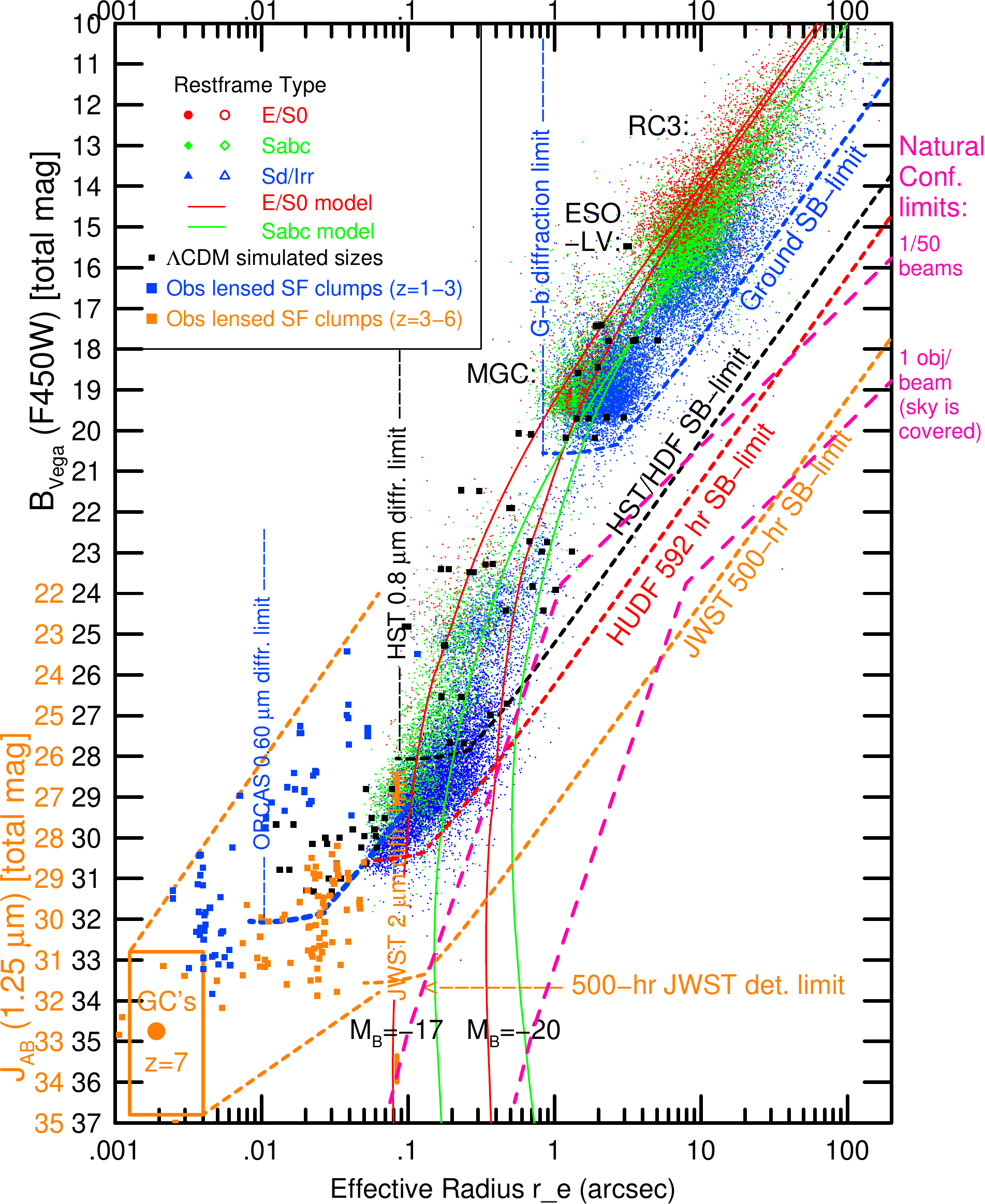}
}

\vspace*{+0.00cm}
\n \caption{
Galaxy sizes \re vs. B$_{\rm Vega}$ or \JAB-magnitude from the RC3 to the HUDF
limit. Short dashed lines indicate survey-limit wedges for the HDF (black),
HUDF (red; in \iAB), and JWST (orange): diffraction limits are vertical,
point-source sensitivity limits horizontal, and SB-sensitivity limits have
slanted slopes = +5 mag/dex. Broken long-dashed pink lines indicate the natural
confusion limit (at the level of 1 detected object per 50 ``beams'' or object
area), to the right of which objects would begin to statistically overlap due
to their own sizes and surface densities. Red and green lines indicate the
expected {\it non-evolving} sizes for RC3 elliptical and spiral galaxies at the
listed \MAB-values, respectively. Blue and orange squares indicate SF-clumps
from gravitational lensing samples with intrinsic physical sizes converted to
\re (\arcs), and unmagnified absolute UV-magnitudes converted to B-mag for
z=1--3 or \JAB for z=3--6 (the \JAB-scale is shown approximately offset in
orange for z\cge 3; the \iAB-scale is between B and \JAB\del{;a non-lensed
z\cge 3 CANDELS sample is shown for comparison at \re\cge 0\arcspt 08}). Black
squares indicate galaxy sizes from hierarchical simulations. The orange box
shows where Globular Clusters at z=7 are expected. For details, see \S\
\ref{sec3}. Most galaxies at \JAB\cge 27--28 mag are expected to be smaller
than the HST and JWST diffraction limits (\ie\ r$_{hl}$\cle 0\arcspt 08). About
half the faint SF-clumps to AB\cle 31 mag are expected to be (barely) resolved
at the ORCAS resolution of 0\arcspt01--0\arcspt 02 FWHM. [Figure adapted from
\cite{windhorst_2008}].
}
\label{fig:fig2}
\end{figure*}

\sn For the success of ORCAS galaxy science, we need to be able to accurately
estimate the expected number density of faint compact star-forming objects out
to z\cle 7 and AB\cle 31 mag. To interpret the currently available lensed
samples of SF clumps, we also need to make an estimate of the {\it intrinsic}
object counts anticipated to AB\cle 35-36 mag. The deepest available data to
date are summarized in Fig~\ref{fig:fig1}a-\ref{fig:fig1}b for the HST ACS/WFC
F606W (wide V-band) and WFC/IR F125W (J-band) filters or their ground-based
equivalents. These data came from the panchromatic ground-based GAMA survey
\citep[which covers AB\cle 18 mag;][]{driver_2010}, the panchromatic HST WFC ERS
survey \citep[17\cle AB\cle 26.5 mag;][]{windhorst_2011}, and the panchromatic
HUDF \citep[22\cle AB\cle 30 mag;][]{beckwith_2006, driver_2016}, and references
therein. The HUDF/XDF limits are indicated by the green labels in the top right
corner of Fig~\ref{fig:fig1}a-\ref{fig:fig1}b. Orange labels indicate the
anticipated JWST Webb Medium Deep Field (WMDF) and UltraDeep Field (WUDF)
survey limits, while red indicates a Webb UltraDeep Frontier Field (WUDFF)
survey limit if pointed at a gravitationally lensing Frontier Field cluster.
The 5$\sigma$ point source detection limits for each of these surveys are
indicated in Fig.~\ref{fig:fig2}, and for both HST and JWST assume an
effective PSF width of 0\arcspt 08 FWHM \citep{windhorst_2008}. Blue labels
indicate the anticipated 5$\sigma$ point source sensitivity limit of AB\cle 31
mag of unlensed objects for ORCAS+Keck with an image PSF with 0\arcspt
01--0\arcspt 02 FWHM at 0.5--1.2 \mum\ wavelength. If ORCAS were to frequently
monitor the best gravitational lensing clusters, we may detect compact sources
intrinsically as faint as AB\cle 35--36 mag when lensed. For an HST and 



\n


\n JWST PSF with FWHM\cle 0\arcspt 08, the depth increase from WUDF to WUDFF is
about 2--3 mag, given the larger unlensed SF-object sizes ($\sim$0\arcspt
005--0\arcspt 080) sampled, while for ORCAS these magnifications could be 
$\sim$3--4 mag for the anticipated {\it smaller} unlensed SF-clump sizes
($\sim$0\arcspt 001--0\arcspt 080) that it may sample (see Fig.~\ref{fig:fig2}).

The observed panchromatic (0.2--1.6 \mum) galaxy counts attain a converging
slope ($\alpha$ $<$ 0.40) for the general flux range of AB$\simeq$17--25 mag
\citep{windhorst_2011, driver_2016}. These counts were fit with models that 
include luminosity + density evolution, as indicated by the four curves in
Fig~\ref{fig:fig1}a-\ref{fig:fig1}b. Some of these models fit the panchromatic
counts remarkably well for 10\cle AB\cle 30 mag. We use the differential count
slope as a function of wavelength \citep{windhorst_2011} to extrapolate the
observed counts to the 31\cle AB\cle 35 mag range. At brighter fluxes, the
0.60--1.25 \mum\ count-slopes are 0.30--0.26 mag/dex, respectively, but for
AB\cge 30 mag we adopt extrapolations with count slopes of approximately
0.15--0.10 dex/mag, as indicated by the blue and orange dashed lines in the
upper-right corners of Fig~\ref{fig:fig1}a-\ref{fig:fig1}b. The justification
for this extrapolation is that the faint-end slope of the galaxy counts is
dominated by galaxies at the median redshift, which in ultradeep redshift
surveys approaches the peak in the cosmic star-formation diagram at
z$\simeq$1.9 \citep{madau_2014}. At this redshift, the best fit faint-end slope
of the Schechter LF is $\alpha$$\simeq$1.4 in linear flux units
\citep{hathi_2010, finkelstein_2016}, so when converted to a magnitude
count-slope, the faint-end slope of the galaxy counts is 
$\gamma$$\simeq$(1.4--1)/2.5 $\simeq$0.16 dex/mag. It is possible that for
fluxes fainter than AB$\sim$31 mag the LF at z\cge 2 --- and therefore the
observed counts --- may turn over with a slope flatter than observed at
brighter levels, but there are arguments against this too \cite[for a 
discussion, see \eg\ \S 2.3 of][]{windhorst_2018}. The adopted extrapolated
slopes in Fig~\ref{fig:fig1}a-\ref{fig:fig1}b are in line with the trend of the
very  faint-end of the plotted galaxy counts models. In both the 0.60 and 1.25
\mum\ filters, the counts integrate to 5.0$\times$10$^6$ faint star-forming (SF)
clumps per square degree to AB\cle 31 mag. (To go from differential to integral
counts in Fig~\ref{fig:fig1}a-\ref{fig:fig1}b, one needs to multiply the
differential surface density at AB=31 mag by 2$\times$ to get the counts per
1.0-mag bin, and by $\sim$3.5$\times$ to get the total integral counts over all
magnitude bins.) This surface density corresponds to $\sim$0.4 SF-object per
arcsec$^2$ to AB\cle 31 mag, or on average one object in every box of
1.6$\times$1.6\arcs. A \cge 5$\times$5\arcs\ FOV of the ORCAS IFU may therefore
provide just enough compact SF clumps to do {it relative} m.a.s.-astrometry as
needed in, \eg\ \S\ \ref{sec4}. 



\vspace*{-0.00cm}
\hspace*{-0.00cm}
\n


\section{The Size Distribution of Faint Star-Forming Clumps to \re\cge
0\arcspt 01 for ORCAS}\label{sec3} 

\sn The median sizes of faint galaxies decline steadily towards higher redshifts
and also towards fainter magnitudes, as shown in Fig.~\ref{fig:fig2}. Red, green
and blue dots show early-type, spiral, and irregular/SF galaxies respectively
\citep[\eg][]{ode96, coh03}. Galaxy structural classification needs to be as
much as possible done at rest-frame wavelengths longwards of the Balmer break
at high redshifts too avoid caveats from the morphological K-correction
\citep[\eg][]{giav96, ode02, win02, tay07}. Red and green lines show the best
fit regression for local galaxies and its extrapolation at fixed \MAB-values to
fainter magnitudes. 

The HST/WFCP2 Hubble Deep Field \citep{williams_1996} and the HST/ACS Hubble
Ultra Deep Field \cite{beckwith_2006} showed that high redshift galaxies are
intrinsically very small with typical sizes of $r_{\rm hl}$$\simeq$ 0\arcspt 12
or 0.7--0.9 kpc at z$\simeq$4--6, and sample correspondingly fainter absolute
AB-magnitudes. The unique combination of these ground-based and HST surveys
shows that the apparent galaxy sizes decline steadily from the RC3 to the HUDF
limits \citep[][and Fig.~\ref{fig:fig2} here]{windhorst_2008}. Most galaxies at
\JAB\cge 28 mag are thus likely unresolved at r$_{hl}$\cle 0\arcspt 1 FWHM, as
suggested by galaxy sizes from hierarchical simulations \citep[black squares in
Fig.~\ref{fig:fig2};][]{kaw04}.

SB and other selection effects in these surveys are significant. For each
survey, the diffraction limit for point sources is shown as vertical dashed
line with the survey indicated, while the nearly horizontal line of the same
color indicates for each survey the corresponding $\sim$5$\sigma$ point-source
sensitivity, and the slanted dashed line (with a of slope 5 mag/dex) indicates
that survey's corresponding SB-sensitivity. That is, each survey cannot detect
objects outside this wedge-shaped area. The pink lines indicate the natural
confusion limit discussed in \citet{windhorst_2008}, that were derived from the
(assumed broken power-law) counts in Fig.~\ref{fig:fig1}a--\ref{fig:fig1}b. As
opposed to the instrumental confusion limit, which is determined by the FWHM of
the PSF in each survey, the natural confusion lines indicate the region where
galaxies would be large enough that their effective area, $\pi$\re$^2$ or ``the
galaxy beam'', would occupy more than 1/50 of the total survey area, thereby
limiting the ability of source detection and deblending algorithms to provide
complete catalogs of overlapping objects. This is primarily visible for
galaxies in the HDF and HUDF for 24\cle B\cle 28 mag and 0\arcspt 4\cle \re\cle
0\arcspt 8, where samples become incomplete as they are no longer bunching up
against the SB-selection lines. Natural confusion is expected to become more
significant for JWST surveys when they are pushed to fainter than
AB$\simeq$30--31 mag. 

Extensive recent studies with HST of several of the best lensing clusters have
resulted in many SF clumps at z$\simeq$1--6.6 that are observed close to the
critical curves, where they appear highly gravitationally stretched and highly
magnified in their total flux \citep[\eg][]{lotz_2017, johnson_2017a,
johnson_2017b, vanzella_2021}. Of particularly importance are the VLT MUSE
spectra and redshifts that have been obtained for many of these SF clumps
\citep{vanzella_2019, vanzella_2021}, which are shown in Fig.~\ref{fig:fig2} 
as the blue (z$\simeq$1--3) and orange (z$\simeq$3--6.6) squares at their {\it
intrinsic} (\ie\ unlensed) physical sizes and {\it unmagnified} absolute
magnitudes (\ie\ their observed \MAB-values after dividing by their lensing
magnification). In Fig.~\ref{fig:fig2} their unlensed physical sizes were
converted to \re in arcsec, and their unmagnified \MUV-values were converted to
B- or \JAB-magnitudes at the corresponding redshifts in $\Lambda$CDM
cosmology. Volume completeness is always hard to estimate even for these best
available gravitational lensing surveys with faint object spectroscopy, but at
least these objects show up in significant numbers in these surveys, and they
populate the {\it unmagnified flux range} of 24\cle AB\cle 34 mag, and the
{\it intrinsic, unlensed size range} of 0\arcspt 001\cle \re\cle 0\arcspt 08 in
Fig.~\ref{fig:fig2}. About half of these SF clumps are expected to be visible
down to the ORCAS diffraction limit, while the other half will be slightly
resolved, but still mostly above the ORCAS SB-limits. A \cge 5$\times$5\arcs\
ORCAS FOV may therefore provide just enough compact SF clumps to do relative
(sub-)m.a.s.-astrometry, depending on the S/R-ratio achieved, which is needed
in \S\ \ref{sec4}. 

Natural confusion is expected to be less important for ORCAS+Keck, since the
sampled unlensed SF-clump sizes from the HST gravitational lensing samples are
much smaller than the HST diffraction limit. Yet is it possible that a number
of larger SF clumps will fall below the ORCAS SB-limits, and only more
hierarchical simulations (black squares) and deeper ORCAS observations will be
able to assess this more precisely.

\section{Monitoring Caustic Transits of Early Stars with ORCAS}\label{sec4} 

\sn Cluster caustic transits can occur when a compact restframe UV source
transits a caustic due to the transverse cluster motion in the sky, or perhaps
due to significant velocity substructure in the cluster, and have the great
potential of magnifying such compact objects temporarily by factors of
$\mu$$\simeq$10$^3$--10$^5$ \citep[\eg][]{miralda_escude_1991}. This is because:
(1) the clusters and their substructures may have transverse motions as high as
\vT\cle 1000 \kms, (2) stars at z$\simeq$1--7 (including population III stars at
z\cge 7) have radii R$\simeq$1--10 \Ro, and (3) in the source plane the main
caustic magnification goes as:\ $\mu$$\simeq$10 $\cdot$ (d$_{\scriptsize
caustic}$/'')$^{-1/2}$, where d$_{\scriptsize caustic}$ is the distance of the
star to the cluster caustic in arcsec. 



\vspace*{-0.00cm} 
\n\begin{figure*}[!hptb] 
\n\cl{ 
\vspace*{-0.00cm} 
\includegraphics[width=1.00\txw,angle=0]{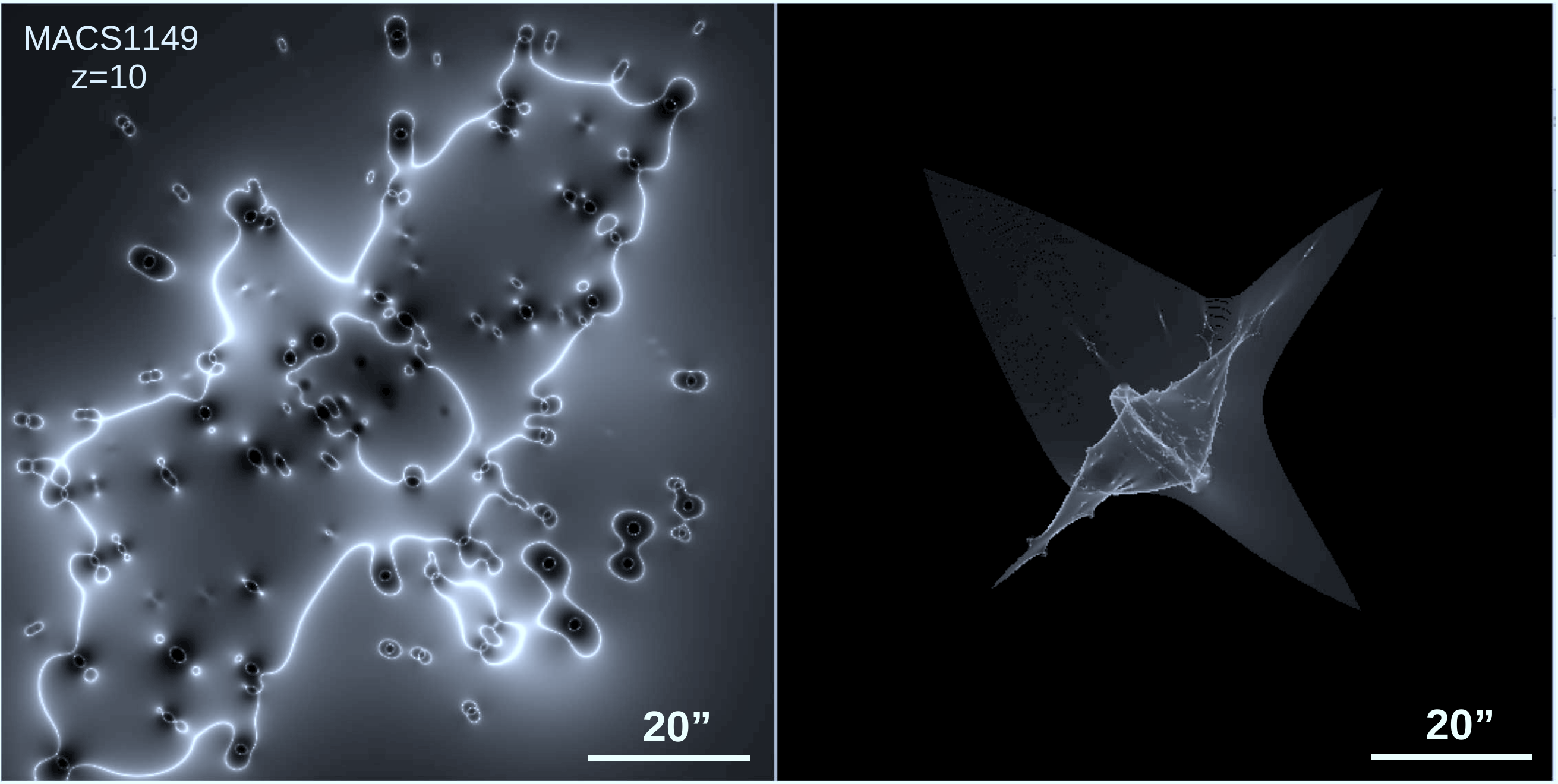} 
} 
\centering{}
\vspace*{-0.300cm} 
\caption{
[LEFT] Example of the lensing magnification map for galaxy cluster MACS
J1149.5+2223 at z$\simeq$0.54 and a background source at z=10 \citep[\eg][and
references therein]{lotz_2017}. Light from the cluster galaxies is not shown to
avoid overcrowding, but can be found in those papers. The white areas mark the
critical curves, where maximum lensing magnification is observed from this
cluster for a background source with half-light radius \rhl\cle 0\arcspt 5 at
z=10. The lightest regions have the highest magnification ($\mu$\cge 10--20),
while the darkest regions are areas of low magnification ($\mu\simeq$1 or even
$\mu$\cle 1) around the cluster member galaxies. 
[RIGHT] Caustic map produced by the cluster mass model for a background source
at z=10. This is the location where a point source at z=10 produces maximum
magnification. The total length of the cluster caustics is L$\simeq$100\arcs
when estimating caustic transit probabilities \citep[for details, see \S\
\ref{sec4} and][]{windhorst_2018}.
} 
\label{fig:fig3}
\end{figure*} 


\n This is illustrated in Fig.~\ref{fig:fig3} as reproduced from
\citet{windhorst_2018}. Since stars at z\cge 7, including Pop III stars, are of
order $\sim$10$^{-11}$ arcsec across at z$\simeq$1--17, such caustic transits
could temporarily boost the brightness of a very compact object by
$\mu$$\simeq$7.5--12.5 mag, which may render it observable by JWST
\citep[\eg][]{windhorst_2018} and also ORCAS+Keck. The best lensing clusters
are typically at z$\simeq$0.3--0.5, and are by selection the most massive,
largely virialized clusters. Lensing clusters with some significant velocity
substructure are preferred, since they tend to have more significant
transverse motions that increase the likelihood of caustic transits. 

In the absence of microlensing by faint stars in the cluster IntraCluster
Light (ICL), these caustic transits may boost the apparent magnitude of these
stars by $\mu$$\simeq$7.5--12.5 mag for several months. This has been observed 
with HST for a number of hot (OB-type) stars at z$\simeq$1--1.5
\citep{kelly_2018, diego_2018, rodney_2018, venumadhav_2017, chen_2019,
kaurov_2019}. \citet{windhorst_2018} calculated the frequency of such events
from both MESA models for Pop III stars and multicolor accretion disks for
stellar mass black holes at z$\simeq$7--17. Both will have roughly the same
radii (R$\simeq$1--100 \Ro) and effective temperatures (\Teff$\sim$50,000--100,000
K), since they will radiate close to the Eddington limit, and therefore they
will have similar rest-frame UV SB. (The only difference is that Pop III stars
never get much hotter than 105,000 K, while stellar mass black hole accretion
disks will also radiate in X-rays when fed from lower mass companion stars in
their AGB stage). Microlensing by faint foreground stars in the cluster ICL
would dilute the macrolensed signal across the main caustic somewhat, but could
also spread it out over more peaks over a longer period of time
\citep[][]{diego_2018}. The resulting expectation is that JWST may observe such
events at the rate of up to $\sim$0.3 per cluster per year if the best lensing
clusters are monitored a few times each year with JWST NIRCam
\citep{windhorst_2018}. 

While the ORCAS FOV is too small for a blind survey of caustic transits at
z\cge 1, it is of particular interest to follow up with ORCAS on caustic
transits of individual stars in SF clumps at z\cge 1--2 that have been detected
with HST, and on caustic transits that may be detected with JWST at
z$\simeq$6--17 at extreme magnifications ($\mu$\cge 10$^3$--10$^5$) for the
first stars and their stellar mass black hole accretion disks. The ability for
ORCAS to monitor such objects for decades across the (micro-)caustics provides
a unique opportunity to obtain a statistical census of individual stars at
cosmological distances. For instance, one could use different ORCAS epochs to
precisely estimate the centroid position of a lensed star that is very close to
a caustic. Assuming the two counter images of the star would form an unresolved
duplet with a separation of less than the ORCAS resolution, microlensing in
each of the two counter images could make the centroid of the observed image
shift from epoch to epoch \citep[\eg][]{diego_2019}, which ORCAS could monitor
at high precision. This then would add a powerful time-domain constraint to 
gravitational lensing models, in addition to the constraints provided by very
deep high-resolution imaging. 

\DELETED{\sn This work was supported by HST grants from STScI, which is
operated by AURA for NASA under contract NAS 5-26555, and by NASA JWST grant
NAG 5-12460.}


\section{Summary of Science Goals and ORCAS Requirements}\label{sec5} 

\sn Here we summarize the ORCAS science goals on faint SF-clumps and their
implications for the ORCAS Requirements Matrix as following. Note that Science
Goal 1+2 may be achieved from other random ORCAS imaging of very faint
foreground targets, such as solar system KBO's:

\bn {\bf Science Goal 1}:\ Constrain the number densities of the faintest
SF-clumps at z$\simeq$1--7. ORCAS will address how galaxies assemble from
smaller clumps to stable disks by measuring ages, metallicities, and gradients
of clumps within galaxies.

\mn {\bf Requirements 1}:\

\sn \bul Deep images to AB\cle 31 mag for point sources in a few hours,
necessary to sample SF-clumps with a surface density of 5.0$\times$10$^6$ per
square degree. 

\sn \bul An \cge 5$\times$5\arcs\ FOV (that includes IFU capabilities), which 
at 5.0$\times$10$^6$ objects per square degree will contain $\sim$10 faint SF
clumps. This is a minimum needed to do {\it relative} (sub-)m.a.s.-astrometry,
depending on the S/R-ratio achieved, and anticipating that most objects will be
compact enough to auto-correlate their images to get the best possible relative
astrometric positions. 

\begin{figure*}[!hptb]
\n\cl{
\includegraphics[width=0.800\txw]{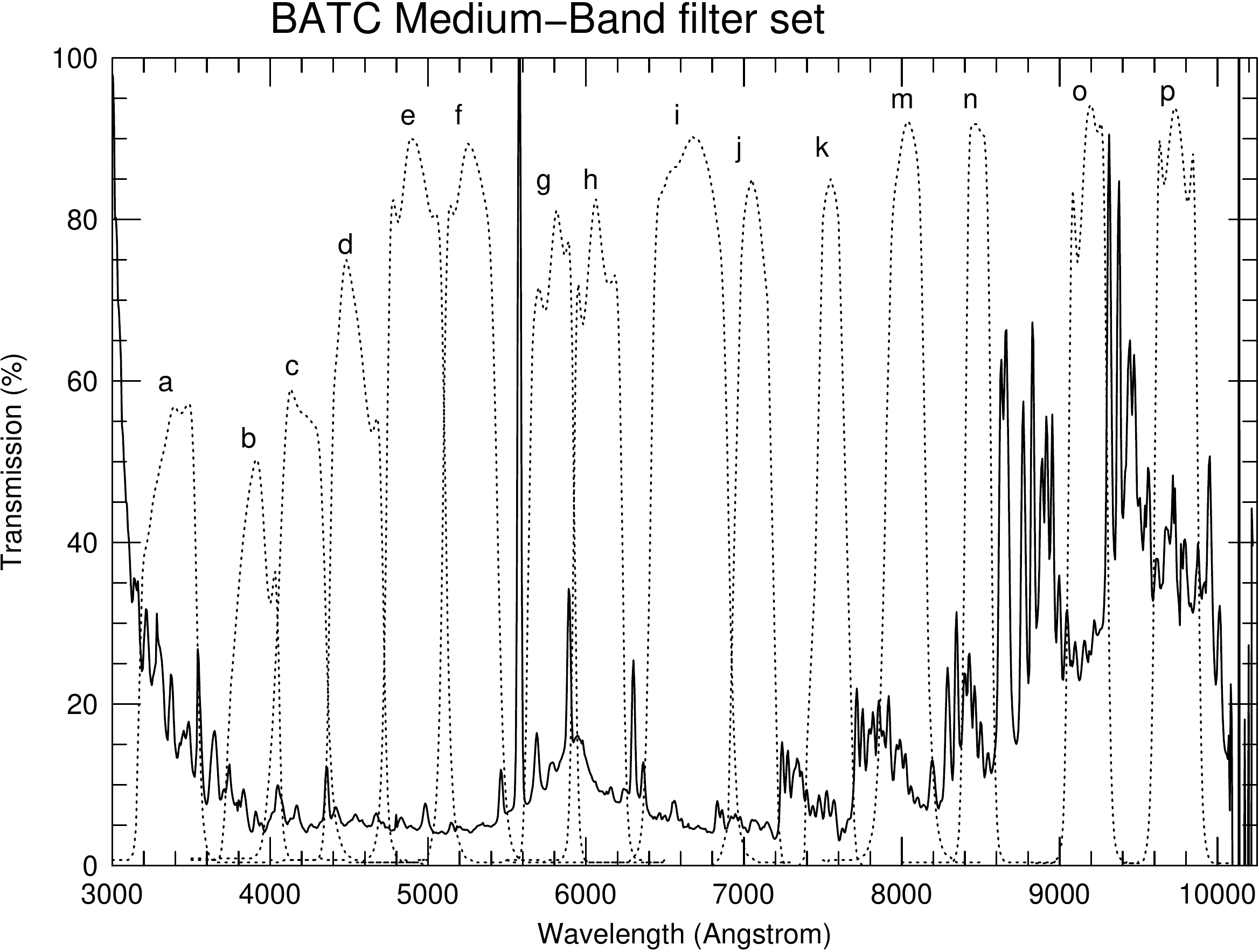}
}

\vspace*{+0.00cm}
\n \caption{
An example of medium-band filters that avoid most of the brightest
night-skylines affecting ground-based observations \citep{yan_2000}. ORCAS will
likely mostly target very faint objects, possibly in a standard broad-band
filter set like ugriz+YJHK. More than half of these objects may be slightly
resolved at ORCAS' spatial resolution of 0\arcspt01--0\arcspt 02 FWHM
(Fig.\ref{fig:fig2}). Hence, to maximize the SB-sensitivity for the very
faintest slightly resolved objects, ORCAS could consider replacing some of the
broad-band filters that include the brightest night-sky lines with
``notch-filters'' that essentially suppress these lines. E.g., one could
replace the V-band or F606W filter with a notch-filter that consist of filter
$f$+$g$ or $g$+$h$ here. This would permit imaging to lower SB-levels
\citep[\eg][]{shang_1998}.
}
\label{fig:fig4}
\end{figure*}

\sn \bul Wavelength coverage ideally at 0.3--2.2 \mum, but at minimum 0.5--1.2
\mum. Standard ugriz+YJHK filter set with potential modifications suggested
below and in Fig.~\ref{fig:fig4}, to get photometric redshift estimates before
IFU spectroscopy is attempted. IFU follow-up on selected targets will be needed.

\sn \bul At minimum 10 ORCAS fields would be needed to start a census for
$\sim$100 of these faint SF clumps. A long term goal should be to get at least
100 ORCAS fields to get a more accurate assessment of the redshift, luminosity
and size distribution from $\sim$1000 SF clumps. 

\bn {\bf Science Goal 2}:\ Constrain the physical sizes of the faintest
SF-clumps at z$\simeq$1--7. Anticipated typical angular sizes at z$\simeq$1--7
are \re$\simeq$1--80 m.a.s. to AB\cle 31 mag. About half of these SF clumps
will be below the ORCAS diffraction limit, and the other half will be slightly
resolved, but still mostly above the ORCAS surface brightness (SB) limits. 

\mn {\bf Requirements 2}:\ 

\sn \bul Spatial resolution of $\sim$0\arcspt 01--0\arcspt 02 FWHM, with good
Strehl ratios.

\sn \bul If SB-sensitivity for the larger SF-clumps becomes an issue, ORCAS
should consider some ``notch-filters'', as shown in Fig.~\ref{fig:fig4}.

\bn {\bf Science Goal 3}:\ Follow up with ORCAS on caustic transits of
individual stars in SF clumps at z\cge 1--2 that have been detected with HST,
and those that may be detected with JWST at z$\simeq$6--17 at extreme
magnifications ($\mu$\cge 10$^3$--10$^5$) for the first stars and their stellar
mass black hole accretion disks. 

\mn {\bf Requirements 3}:\

\sn \bul Deep images to AB\cle 31 mag for point sources. Unmagnified
magnitudes (\ie\ their observed lensed fluxes after dividing by their lensing
magnification) may be as faint as AB\cle 35--36 mag. 

\sn \bul Spatial resolution of $\sim$0\arcspt 01--0\arcspt 02 FWHM. Unlensed
sizes may be smaller than 1--10 m.a.s. when searching with ORCAS around the
critical curves of the best lensing galaxy clusters imaged with HST and JWST. 

\sn \bul {\it Relative} (sub-)m.a.s.-astrometry will be needed to monitor
potential parity changes of lensed sources when they go across a caustic, and
therefore may change shape or apparent position at high redshift (Note: this
is not a true proper 


\vspace*{-0.00cm}
\hspace*{-0.00cm}
\n


\n motion, but a light-path change of the lensed source in the gravitational
landscape of the lensing cluster when the source goes across a caustic.

\sn \bul Monitor caustic-transiting stars for decades across the
(micro-)caustics to obtain a statistical census of individual stars at
cosmological distances, and the microlensing stellar population in the
foreground galaxy cluster ICL. 

\sn \bul Preimaging with HST or follow-up imaging with JWST of ORCAS targets 
may be needed to identify the best possible candidates for caustic transits.

\bibliographystyle{aasjournal}



\hspace*{-0.00cm}

\end{document}